\documentclass[11pt,a4paper,oneside]{article}
\usepackage{amsmath, amssymb,amsthm}
\usepackage{graphicx,subfigure,hyperref}
\pdfoutput=1
\hypersetup{pdfauthor={Ingo Scholtes and Claudio Juan Tessone},pdftitle={Organic Design of Massively Distributed Systems: A Complex Networks Perspective}}

\begin{document}
\title{Organic Design of Massively Distributed Systems: \\ A Complex Networks Perspective}

\author{Ingo Scholtes and Claudio Juan Tessone \\
  ETH Z\"{u}rich\\
  Chair of Systems Design\\
  CH-8032 Z\"{u}rich\\
  Switzerland\\
  \texttt{\{ischoltes,tessonec\}@ethz.ch}}
\date{December 18th, 2011}
\maketitle

\abstract{\textbf{Abstract:} The vision of \emph{Organic Computing} addresses challenges that arise in the design of future information systems that are comprised of numerous, heterogeneous, resource-constrained and error-prone components or devices. Here, the notion \emph{organic} particularly highlights the idea that, in order to be manageable, such systems should exhibit self-organization, self-adaptation and self-healing characteristics similar to those of biological systems. In recent years, the principles underlying many of the interesting characteristics of natural systems have been investigated from the perspective of complex systems science, particularly using the conceptual framework of statistical physics and statistical mechanics. In this article, we review some of the interesting relations between statistical physics and networked systems and discuss applications in the engineering of organic networked computing systems with predictable, quantifiable and controllable self-* properties.}

\section{Introduction}
\label{sec:Introduction}

Networked computing systems are becoming increasingly large, complex and --at the same time-- important for our everyday lives. Many of the services and applications we rely on every day are now being collaboratively provided by thousands or even millions of machines in large-scale Peer-to-Peer systems or data centers. Sustaining the robustness and manageability of such systems is a challenging task. However the associated challenges are likely to aggravate significantly in the future. Due to the ongoing miniaturization of networking technologies, its decreasing price as well as the proliferation of mobile and embedded computing equipment, scenarios in which several billion devices are connected to global-scale information systems come into reach. Promising aspects of the coalescence of the virtual and physical world that results from the increasing incorporation of communication technology into everyday objects, as well as the associated technical and societal challenges have been highlighted in the visions of \emph{Ubiquitous Computing} or the \emph{Internet of Things} \cite{Weiser1991a,Mattern2010}.

Clearly, building services and applications in an environment of increasingly numerous, heterogeneous, dynamic and error-prone communication devices poses enormous technical challenges in terms of scalability, complexity, efficiency, manageability and robustness. It is frequently argued that, in order to cope with these challenges, computing technologies need to adopt some of the remarkable self-organization, self-adaptation and self-healing qualities of biological systems. In recent years, facilitated by advances in the study of principles underlying self-organization mechanisms as well as the massively increasing complexity of technical infrastructure, the associated vision of \emph{Organic Computing} is gaining momentum. The development of technologies suitable for systems with organic qualities is likely to go in hand with a cutback of highly sophisticated algorithmic schemes and deterministically organized, rigid communication and data structures. Instead, adopting a more heterodox approach that utilizes some of the simple mechanisms that underlie self-organization, adaptivity, robustness and resilience in natural systems appears reasonable. 

Driven mainly by the availability of massive data sets, during the last decade these principles have been studied in a variety of different contexts, including dispair fields such as biology, physics, computer science, economics and sociology. The resulting interdisciplinary strand of research is typically subsumed under the name of \emph{complex systems science} and here we argue that it offers a promising and quickly evolving methodological framework for the modeling, design, control and analysis of \emph{organic computing systems}. Providing a set of tools and abstractions that allow to analyze the collective properties of systems comprised of a large number of stochastic, interacting elements, complex systems science addresses one of the key questions emerging in the scenario described above: How can we predict, monitor and control the structure and dynamics of massively distributed systems based on maximally simple distributed mechanisms?

In this article we address some aspects of this question that arise in the management of scalable, robust and adaptive overlay topologies in massively distributed networked computing systems. For this, we will adopt the perspective of statistical mechanics on the emergence of complex structures and collective dynamics in networked systems -- an area that has been particularly active and successful during the last decade. In section \ref{sec:Networks} we first summarize structured and unstructured approaches to the management of overlay topologies in large-scale systems. Here we further review the relevance of random graph theory in the design of unstructured systems and introduce some interesting relations between the study of statistical mechanics and complex networks as well as in the modeling of dynamical processes unfolding within them. In section \ref{sec:Organic} we then discuss how --based on these relations-- abstractions and notions from statistical mechanics and statistical physics can be used in the design of distributed management schemes for organic overlay networks. In section \ref{sec:Conclusion} we finally discuss some challenges and opportunities that emerge when using complex systems science in the engineering of organic networked computing systems with predictable, quantifiable and controllable self-* properties.

\section{Overlay Topologies, Random Graphs and Complex Networks}
\label{sec:Networks}

Overlay networks --which define virtual connections on top of physical communication infrastructures-- are becoming an increasingly important abstraction. As argued for example in \cite{Steinmetz2005,Waldhorst2010}, the possibility to define communication topologies and protocols at the application-layer without having to go through the painstaking process of making a --potentially globally-- coordinated change of existing protocols, standards and communication infrastructure is an important factor for the proliferation of novel services on the Internet as well as in large-scale data centers. The research of scalable and robust overlay topologies as well as efficient distributed algorithms providing core functionality like search, routing and content dissemination has received a lot of attention during the last couple of years.

Most of this research has been done in the context of large-scale Peer-to-Peer systems, which are now increasingly being used for the cost-efficient distribution of large amounts of data for example by means of the BitTorrent protocol, the provision of video-telephony services like Skype or even to face the challenges emerging in large-scale scientific setups like the Large Hadron Collider\cite{Scholtes2008c}. Regarding the overlay networks employed by these systems, one usually distinguishes \emph{structured} and \emph{unstructured} approaches. Most of the currently deployed systems belong to the former category. In such structured systems, virtual connections between participating machines are typically created in a globally consistent way to construct a predetermined topology. While this allows for the development of highly efficient algorithms for distributed search, routing or information dissemination, the fact that this fine-tuned topology has to be maintained at all times in the face of highly dynamic constituents is a major drawback. Reconsidering the scenario outlined in section \ref{sec:Introduction}, correctly and efficiently maintaining fine-tuned structures will entail massive complexities due to the excessive fluctuation of participating devices and the associated concurrency. In fact, for the popular distributed hash table overlay \emph{Chord} it has been argued in \cite{Balakrishnan2003} that in settings with very large numbers of highly dynamic participants, the communication overhead imposed by mere topology maintenance and management schemes exceeds the cost for actual lookup and data transfer operations and thus dominates performance. Here it has further been argued that correctly designing, implementing and debugging topology maintenance operations poses a huge challenge due to the massive concurrency that is introduced by failing or joining machines. These problems of structured overlay topologies are well-known in literature and they cast their usability in future scenarios like the one laid out in section \ref{sec:Introduction} into doubt. Hence, alternate approaches to deal with massive-scale, dynamic systems are increasingly being studied. 

\subsection{Unstructured Topologies and Random Graph Theory}
\label{sec:Networks:RG}

A straight-forward idea is the use of unstructured topologies in which virtual connections between nodes can be created in a much simpler, uncoordinated fashion as long as they still allow all participating machines to communicate. While this significantly reduces the topology management overhead, it necessitates the use of probabilistic algorithms e.g. for distributed search or routing that make no --or at least less specific assumptions-- about the structure of the communication networks or the placement of data items. Such schemes are inevitably less efficient than those tailored to a particular network structure. They are however significantly simpler to implement and allow for larger degrees of freedom in terms of adapting the network structure to operational conditions.

In terms of modeling their performance and robustness, most unstructured approaches to the management of overlay topologies rely --either explicitly or implicitly-- on fundamental results from the field of random graph theory, which was established more than 50 years ago \cite{Erdos1959}. In order to be able to appreciate the analogies between the management of large, dynamic networked systems, statistical mechanics and thermodynamics, we first briefly recall one of the basic models of random graph theory. This so-called $G(n,p)$ model defines a probability space that contains all possible graphs or networks\footnote{Throughout this article, we will use the terms \emph{graph} and \emph{network} interchangeably.} $G$ with $n$ nodes. Assuming that edges between pairs of nodes are being generated by a stochastic process with uniform probability $p$, the $G(n,p)$ model assigns each network $G=(V,E)$ with $n=|V|$ nodes and $m=|E|$ edges the same probability

\begin{equation*}
	P_G(n,p) = p^m \cdot \left(1-p\right)^{n (n-1)/2-m}
\end{equation*}
This simple stochastic model for networks has been used in the modeling of a variety of structural features of real-world networks. In particular, it can be used to make predictions about the properties of unstructured overlay topologies, if virtual connections are assumed to be created at random with probability $p$ or, alternatively, if an average number of $p\cdot n(n-1)/2$ connections are established between randomly chosen pairs of nodes.

In general, in the study of random network structures one is particularly interested in properties that hold for a subset of network realizations whose probability measure converges to $1$ as the size of the generated networks (in terms of the number of nodes) increases. In this case one says that for a randomly generated topology a property holds \emph{asymptotically almost surely}. A number of interesting results have been derived based on this perspective and an authoritative overview can be found in \cite{Bollobas2001}. In the design of unstructured overlays it is valid to rely on these results if the system is sufficiently large and the convergence of the probability is sufficiently fast. Two well-known examples that are of particular relevance for the design of overlay topologies are results on the critical percolation threshold and the small-world phenomenon. The critical percolation threshold refers to a critical point in the $G(n,p)$ model's parameter $p$ above which the generated networks almost surely contains a connected component that is of the order of the network size. For the $G(n,p)$ model it has been found that connected components of a random graph are with high probability of the order $log(n)$ if $p < 1/n$. For $p > 1/n$ the connected component is of the order $n$ \cite{Erdos1959}\footnote{Interestingly this is a so-called double-jump transition, i.e. for $p=1/n$ the size of the connected component is of the order $n^\frac{2}{3}$}. In practical terms, this result is a crucial prerequisite for the feasibility of unstructured overlay topology management schemes since it tells us that --if at least a certain minimum number of connections is created in a random, uncoordinated fashion-- all machines will be able to directly or indirectly communicate with each other with high probability. A further set of results which are important for overlays with random structures relates the parameter $p$ --or alternatively the average number of randomly created connections-- to the diameter of the resulting topology. It further gives a criterion for the emergence of so-called \emph{small-world topologies} which usually are assumed to have a diameter of the order of the logarithm of the network size. For the $G(n,p)$ model, it has been shown that the diameter is with high probability of order $log(n)/log(np)$ if the average number of links per node is at least $1$. In the design of unstructured topologies this argument is being used to reason about the efficiency of search and routing schemes.

\subsection{Statistical Mechanics of Complex Networks}
\label{sec:Networks:Complex}

As argued in \cite{Cohen1988}, the existence of critical points in the $G(n,p)$ model's parameter $p$ and the associated sudden change of macroscopic network qualities like diameter or connectedness highlights interesting relations to phase transition phenomena in statistical physics, i.e. sudden changes of material properties as aggregate control parameters (like e.g. temperature or pressure) change slightly. In recent years, these analogies to fundamental natural phenomena have been deepened substantially by reframing the study of random graph structures in terms of statistical mechanics and statistical physics (see e.g. \cite{Berg2002,Albert2002, Dorogovtsev2003,Farkas2004,Garlaschelli2008,Reichardt2006}). This perspective is possible since, at its very foundation, statistical mechanics reasons about configuration probabilities of many-particle systems, just like random graph theory reasons about the probabilities of network configurations. Each of these particle configurations --the so-called \emph{microstate}-- fixes the exact positions and energy states of all particles present in a given volume of space at a given temperature and total energy. Based on energy distributions, particle positions and fluctuations induced by temperature, each of these microstates can be assigned a probability. The set of all possible microstates thus defines a probability space which is called, in the language of statistical mechanics, a \emph{statistical ensemble}. The success of statistical mechanics in explaining macroscopic properties of matter based on a statistical picture of microscopic particle dynamics started with the study of a discrete approximation of space and a quantization of energy. Initially this approach was chosen as a rough abstraction to render microstate probabilities accessible by mere combinatorial arguments much in the same way as graph probabilities are accessible in random graph theory.\footnote{Interestingly this abstraction was substantiated by the development of quantum physics roughly 40 years later.}

Based on these similarities, it has been argued for instance in \cite{Garlaschelli2008} how the $G(n,p)$ model of classical random graphs can be reframed in terms of the so-called \emph{grand-canonical ensemble} of many-particle systems residing in thermodynamic equilibrium. In this framework, the study of statistical ensembles with fixed thermodynamic quantities like volume, chemical potential and temperature in statistical mechanics translates to the study of \emph{network ensembles} with fixed aggregate statistics, like for instance a given number of nodes or edges, degree distribution, degree-degree correlations or clustering coefficients. In the resulting statistical ensembles all realizations with the same aggregate statistics (e.g. all networks with a particular degree sequence) are assumed to have equal probabilities. In other words, networks are assumed to be sampled from a probabilistic mechanism which does not differentiate between realizations with the same degree distribution just like the stochastic process underlying the $G(n,p)$ model does not differentiate between two realizations with the same number of edges. In statistical mechanics, this corresponds to an \emph{adiabatic} situation in thermodynamic equilibrium while at the same time the accessible states are being constrained by certain fixed quantities. In the remainder of this article we will thus refer to such probability spaces as \emph{constrained adiabatic ensembles}.

During the last decade such constrained adiabatic ensembles of random network structures have been studied extensively in the fields of \emph{complex systems}, \emph{complex networks} and \emph{statistical mechanics}. A particularly active strand of research in this direction is the study of ensembles with fixed degree sequences, or degree distributions following, for instance, a power-law. This is, the probability that a randomly chosen node in the network has exactly $k$ links, is proportional to $k^{-\gamma}$ for some $\gamma \in [2, \infty)$. As the classical $G(n,p)$ model can be viewed alternatively as an adiabatic ensemble with a fixed Poissonian degree distribution, these studies naturally extend earlier works on random graphs. This is particularly true since power-laws (or at least heavy tailed) degree distributions have been observed for a number of real-world networks \cite{Albert2002}. As such, the associated adiabatic constrained ensemble effectively serves as a null model for these kinds of systems. 

Over the last years, a rich set of results both on the collective properties of such networks, as well as on simple local mechanisms giving rise to such structures have been obtained. Extensive surveys of results based on this statistical physics perspective on complex network structures can be found for instance in \cite{Boccaletti2006,Barrat2008}. Prominent results for the special case of networks with heavy-tailed degree distributions include their resilience against random faults \cite{Cohen2000} and targeted attacks \cite{Cohen2001} or the performance of probabilistic distributed search schemes \cite{Adamic2001}. In the remainder of this article, we argue that such results and --even more importantly-- the methodological framework with which they have been derived are highly relevant for the design of robust networked systems with organic properties.

\subsection{Dynamical Processes in Complex Networks}
\label{sec:Organic:Dynamic}

So far, we have commented on the structural properties emerging in networks being drawn from constrained adiabatic ensembles. For the design of distributed algorithms that need to operate in an efficient and reliable way in large dynamic overlay topologies, it is equally important to have tools at hand that allow to reason about the properties of dynamical processes. Fortunately, during the last couple of years such tools have been developed in random graph theory and complex network science. In order to formalize the problem, a useful representation of a graph is in terms of its adjacency matrix ${\mathbf A}$ where each element is $a_{ij} = 1$ ($a_{ij} = 0$) if the nodes $i$ and $j$ are connected (respectively, disconnected). Then, the {\em spectrum} of such a network is given by the set of $n$ eigenvalues of its adjacency matrix. For the $G(n,p)$ model, it is possible to characterize the spectrum of the networks in the limit of diverging network sizes. In this regime, and if there is a giant cluster that spans the complete network, the probability $p(\lambda)$ of finding an eigenvalue $\lambda^{a}$ in the spectrum follows the so-called {\em semi-circle law} \cite{Mehta1991},
\begin{eqnarray*}
p (\lambda^{a}) =  %
\begin{cases}
  \frac{\sqrt{ 4\, n \, p (1-p) - \left( \lambda^{a} \right) ^2 }}{2\pi\, n \, p (1-p)} & \hbox{if} \,\,|\lambda^{a}| < 2 \sqrt{n\, p(1-p)} \\
  0  & \hbox{if} \,\,  |\lambda^{a}| \geq 2 \sqrt{n\, p(1-p)}.
\end{cases}
\end{eqnarray*}
The bulk of the distribution of eigenvalues is centered around the null eigenvalue, with a characteristic size proportional to $\sqrt n$. However, the largest eigenvalue $\lambda^a_{1}$, is proportional to $p\,n$.

The nature of processes unfolding in networks may give rise to different kind of phenomena. In particular, when the processes can be described by a variable (or a set of variables) evolving continuously in time, one talks about {dynamical processes in networks}. 
In order to this kind of processes, we consider that each node $i$ is endowed with a continuous variable $x_i$ which describes its current dynamical state. Then, the change of its state can be thought to be given by
\begin{equation}
\frac{d}{dt} x_i = f_i(x_i) + C \sum_{j=1}^n \ell_{ij} \cdot h(x_j), \label{eq:dynevol}
\end{equation}
where $f(x_i)$ is a function describing a deterministic change of state of node $i$ given its current state, and $h(x_j)$ is a coupling function given the state of node $j$, and $C$ is the coupling strength. In Equation \ref{eq:dynevol}, $\ell_{ij}$ are the elements of the {\em Laplacian matrix}, ${\mathbf L}$. Such matrix is defined as $\ell_{ij} = - k_i \delta_{ij} + a_{ij}$, where $k_i$ is the degree of node $i$, and $\delta_{ij}$ are the elements of the identity matrix. The Laplacian matrix, receives such name because it naturally extends the Laplacian operation --as used in spatially extended systems, $\nabla^2$-- into a discrete manifold.

Different kinds of collective behavior may appear in when studying dynamical processes in complex networks \cite{Boccaletti2006,Arenas2008}. 
When the different nodes exhibit common dynamics at the global level, it can be said that a synchronized state has emerged. In general, it was shown \cite{Barahona2002} that the synchronized state is stable if $\lambda^l_n/\lambda^l_2 < \beta$, where $\lambda^l_n$ and $\lambda^l_2$ are (respectively) the largest and smallest non-zero eigenvalues of the Laplacian matrix, and $\beta$ is given by the functions describing $f(x)$ and $h(x)$ describing the dynamical properties of the system. Interestingly, this means that there exists a relationship between structural properties of the network (in terms of its eigenvalues) and the dynamical process taking place at the level of nodes as to whether the system is able to synchronize. This is against the typical intuition on synchronization phenomena, stating that if the coupling strength is large enough, the system should exhibit some degree of synchronization.

At a more detailed level, if a synchronized behavior emerges the question of {\em how much time} does the system need to reach a synchronized state usually termed as {\em consensus dynamics}. For arbitrary networks it was shown \cite{Almendral2007} that the consensus time is of the order $T_C = (\ln C - \ln \epsilon /2 ) / \lambda^l_2 $, where $C$ is an integration constant (related to the initial conditions) and $\epsilon$ is the {\em synchronization threshold}: the difference below which the dynamics are assumed to be common. 

Statements on the impact of the Laplacian spectrum play an important role when assessing the properties of synchronization and consensus schemes operating in large-scale networked systems. In overlay networks such models can for instance be used to provide a network-wide synchronous notion of time epochs or protocol cycles. Examples for synchronization protocols which effectively rely on spectral properties of unstructured and semi-structured topologies can be found for instance in \cite{Lucarelli2004,Babaoglu2007,Baldoni2009,Scholtes2010b}. Furthermore, this methodological framework plays an important role for the mitigation of detrimental synchronization phenomena as for instance observed in \cite{Floyd1994}.

Dynamic processes can take place also in a discrete space. Perhaps the simplest, still important, of such processes, is the random walk in a graph. Such process is equivalent to stochastic search, and transport phenomena within the network. In these contexts, how fast can a node receive and spread information over the network in such a random process is very important \cite{Noh2004}. To quantify this, the {\em random-walk centrality} $\mathcal{C}_i$ was introduced which --for node $i$-- is given by,
$$
\mathcal{C}_i = \frac{k_i}{\sum_i k_i} \left( P_{ii}(t) -  \frac{k_i}{\sum_i k_i}\right)^{-1}.
$$ 
Where $P_{ii}(t)$ is the probability that a random walk which started at $i$ at time zero, returns at the same node after $t$ time steps, and can be readily computed by means of the adjacency matrix $\mathbf{A}$.
This index shows that nodes with large centrality receive information faster. In the case of limited bandwidth capacities, it also indicated the fact that such nodes may become overloaded.
Arguments about the relation between spectral properties of networks and the dynamics of random walk processes have been used --again implicitly or explicitly-- in a variety of practical contexts. Being used for multicast communication \cite{Gupta2006}, database replica maintenance \cite{Demers1987}, the computation of network-wide aggregates \cite{Jelasity2005} or random sampling \cite{Zhong2008}, gossiping and random walk protocols are one of the most successful classes of distributed probabilistic schemes in unstructured or semi-structured overlays. By means of the spectral perspective on the equivalent of the Laplacian operator in networks, it is thus possible to study the performance of these schemes for networks drawn from a particular statistical ensemble. This clearly demonstrates the relevance of these techniques for the design of organic networked systems, while at the same time highlighting interesting mathematical relations to fundamental natural phenomena being studied in statistical physics.

\section{Managing Organic Overlays - A Thermodynamic Perspective}
\label{sec:Organic}

From an engineer's perspective, the results on the structural and dynamical properties of complex networks that have been reviewed in section \ref{sec:Networks} may sound appealing. However, one needs to be careful when applying them to actual systems. As mentioned above, statistical ensembles of complex networks should be viewed as mere \emph{null models} for networked systems with complex structures. As such, it is clearly unlikely that they are able to accurately reproduce the structures and properties of highly sophisticated infrastructures like the Internet which are subject to numerous technological constraints, engineering principles and substantial planning. Some of the fallacies and challenges that arise when imprudently applying the complex networks perspective to sophisticated technical infrastructures have been summarized for instance in \cite{Willinger}.

Nevertheless, given that the statistical mechanics perspective on networks is able to reason about structural and dynamical properties that are relevant for the design and operation of networked systems in general and overlay networks in particular, it is reasonable to study how one can use simple stochastic models for complex networks \emph{in a constructive rather than in an explanatory way}. After all, topology management schemes can be explicitly designed along a stochastic model that gives rise to a class of network topologies  whose structural features and dynamical properties are advantageous for a particular setting. By means of distributed probabilistic protocols --like for instance suitably configured random walk sampling or rewiring schemes-- reproducing a stochastic model is often much simpler than designing, implementing and debugging sophisticated distributed algorithms that precisely manage overlay topologies.

One can actually use the perspective of statistical mechanics to contrast this approach with traditional structured and unstructured topology management schemes: From this point of view structured topology management schemes give rise to states of small (statistical) entropy in the sense that only a small subset of all thinkable network realizations are accessible. From the statistical ensemble point of view the associated probability distribution is a delta function that minimizes entropy. At the same time this maximizes the amount of information one has about the detailed structures of the topology. It is this information that can be used to design highly efficient distributed algorithms that utilize the network structure to provide efficient key lookups, routing and information spreading. At the same time, maintenance routines are required to prevent the gradual loss of structure --and thus increase of entropy-- that is due to the dynamics of users and devices, the occurrence of hard- and software failures or communication errors. As such, the computation and communication effort induced by topology maintenance mechanisms can be viewed in analogy to the input of energy that is used by non-equilibrium biological systems to prevent the entropy maximization that is due to the second law of thermodynamics. Analogously, unstructured approaches can be related to states of maximum statistical entropy in which all realizations of a network are equally likely because the topology is constructed in a purely random fashion. In this case, distributed algorithms can use no a priori information about the detailed structures of the network structure, thus often leaving flooding or exhaustive search as the only viable options.

One of the most interesting aspects of complex network science is that it allows to explore --both analytically and in terms of suitable stochastic models-- the interesting middle-ground of \emph{statistically} or \emph{thermodynamically} structured systems with intermediate levels of entropy. The resulting topologies are neither completely random nor completely deterministic. The fact that the emerging network structures are due to the introduction of mere statistical correlations rather than rigid topology management schemes allows for variation and adaptivity that is required to cope with unforeseen situations or changing operational conditions. Still such topologies can exhibit enough statistical structure (like a particular type of degree sequence, a certain clustering coefficient or correlations between data location and network structures) to solve algorithmic tasks much more efficiently than in unstructured systems. For distributed search in Peer-to-Peer systems, it has been shown for instance in \cite{Sarshar2004} how the structure of randomly generated networks with a power law degree distribution can be exploited in order to improve search performance. Similar results have been obtained for routing schemes making use of correlations between node addresses and network structure, particular clustering structures or an embedding in Euclidean or Hyperbolic coordinate spaces \cite{Sandberg2006,Kleinberg2006,Thadakamalla2007,Boguna2008,Papadopoulos2010}.

\subsection{Structure Formation in Overlay Networks}
\label{sec:Organic:Equilibrium}

As mentioned above, considering the question of structure formation and maintenance in overlay networks from the perspective of thermodynamics and statistical mechanics can be insightful. For the design of probabilistic overlay management protocols, thinking in equilibrium and non-equilibrium states provides a number of interesting perspectives: In equilibrium situations, the formation of complex structures in network topologies can be viewed as the result of a stochastic optimization process resulting from the minimization of free energy and the existence of random thermal fluctuations. The reformulation of network models in terms of equilibrium statistical ensembles, allows us to further identify if a networked system is out of equilibrium. Examples for this include systems which undergo phases of growth, in the sense that the average number of nodes and/or edges increases (resembling an influx of mass), or the average resources or resource demands change (resembling a change of chemical potential or free energy). The fact that complex structures have been observed in the overlay of systems like Gnutella (which started out as being unstructured) can in fact be related to growth and a non-equilibrium dynamics of nodes (e.g. in terms of highly skewed session times). Tools from statistical mechanics dealing with such non-equilibrium situations have actually been used to study the self-organized structure formation in said systems \cite{Barabasi1999}.

\begin{figure}[!ht]
  \centering
  \subfigure[Random initial network]{\label{fig:Annealing:a} \includegraphics[width=5cm]{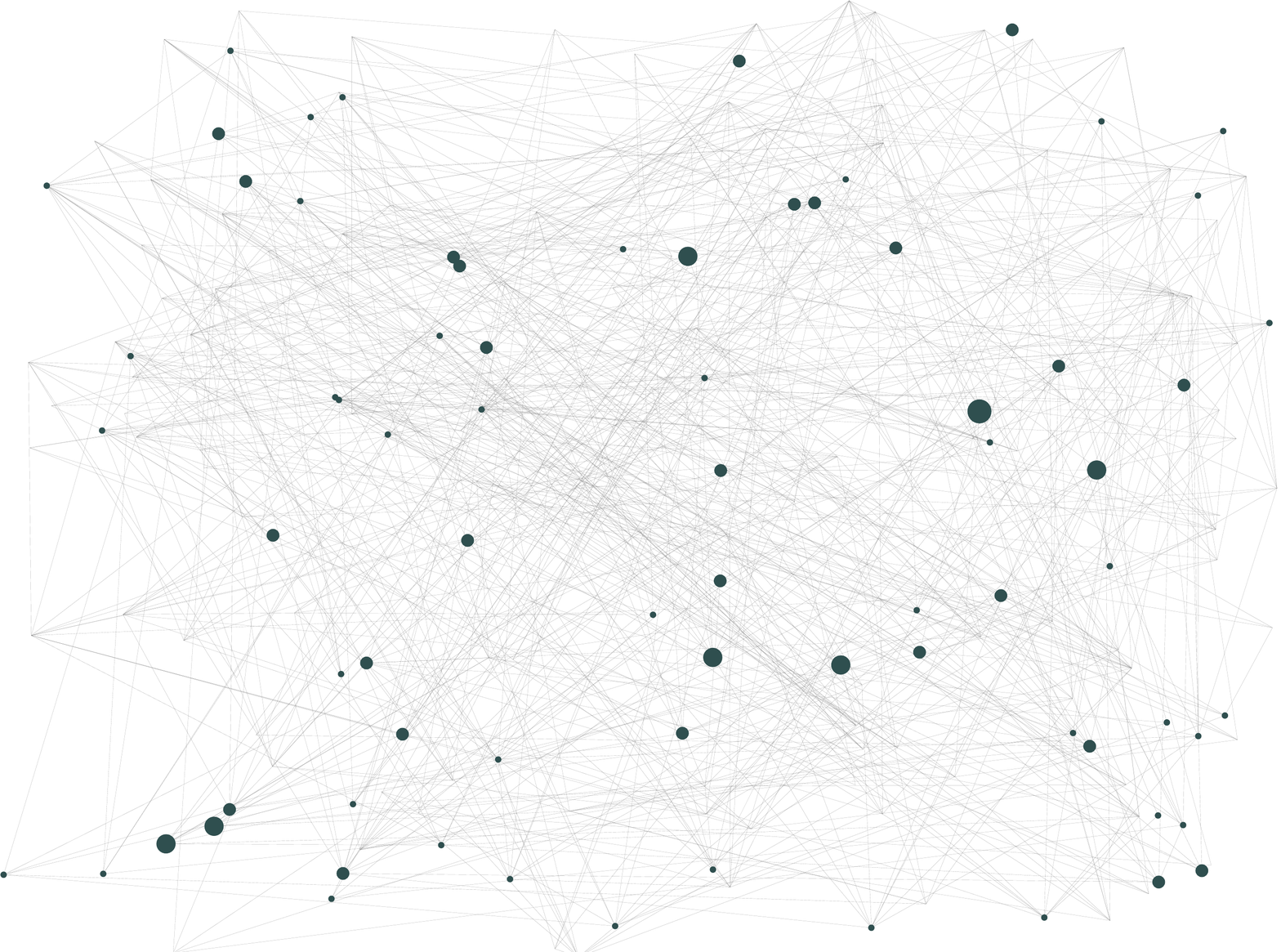}}
  \subfigure[Intermediate state]{\label{fig:Annealing:b} \includegraphics[width=5cm]{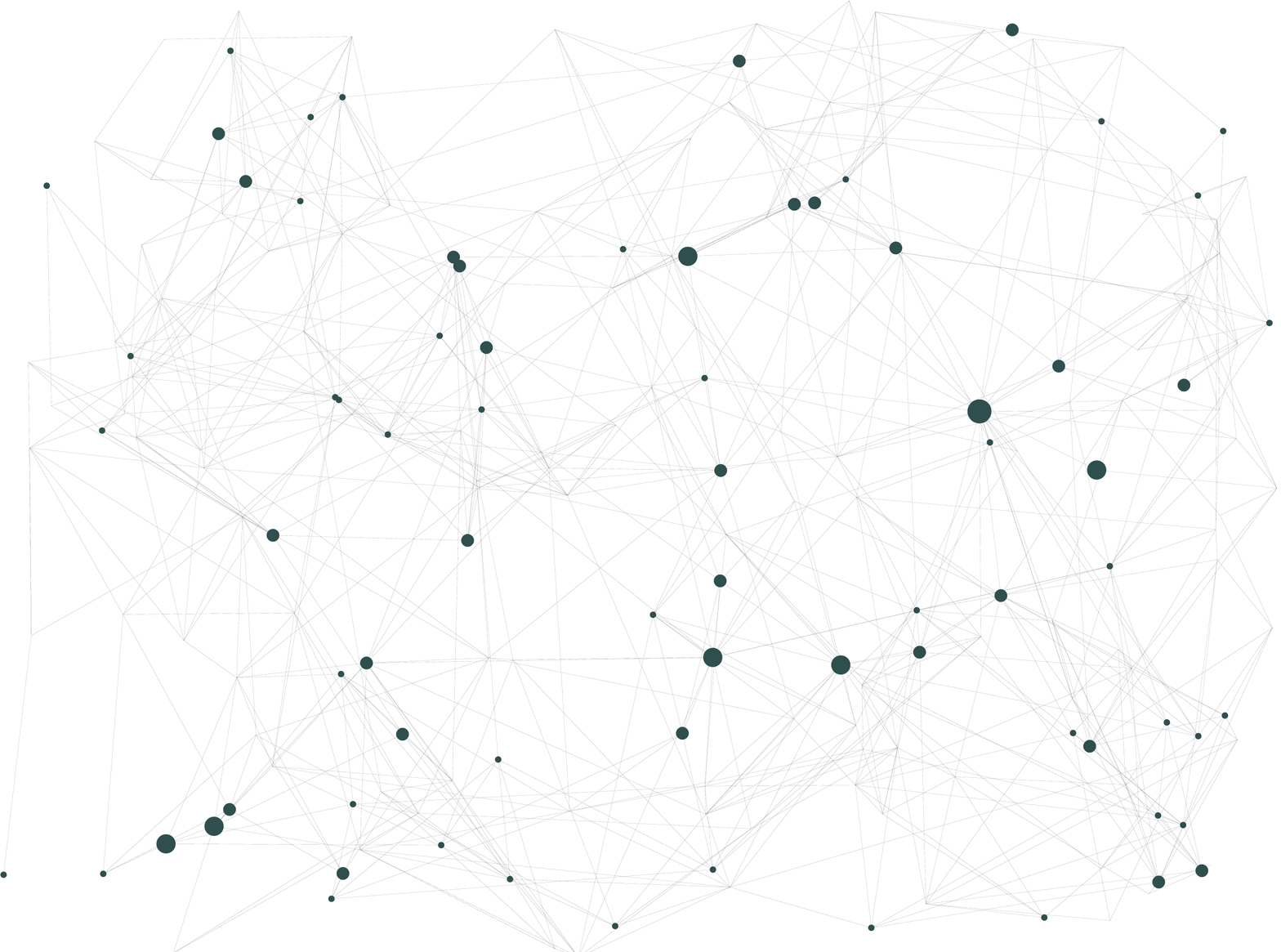}} 
  \subfigure[Equilibrium network]{\label{fig:Annealing:c} \includegraphics[width=5cm]{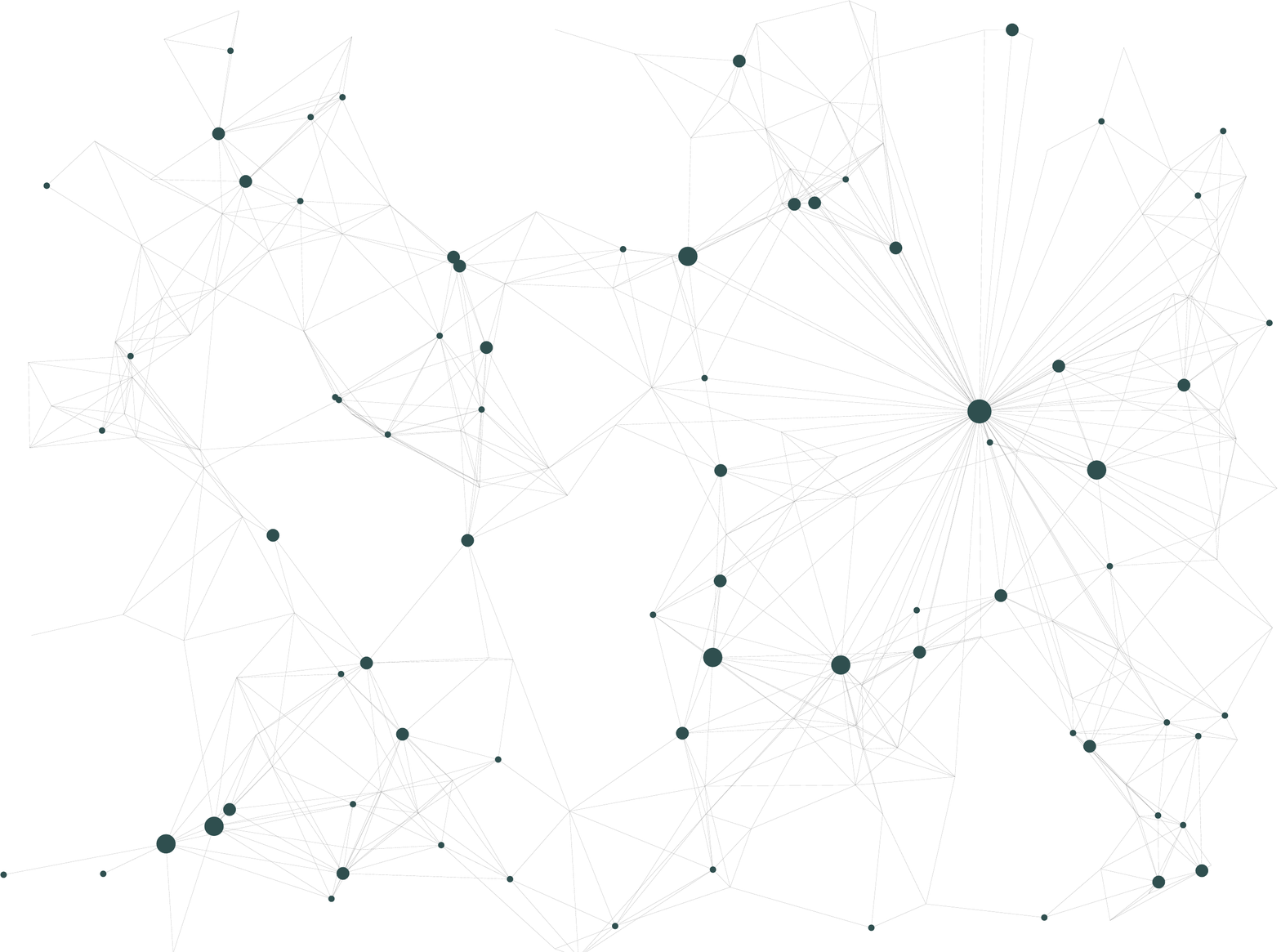}} \qquad
  \subfigure[Change of node capacities]{\label{fig:Annealing:d} \includegraphics[width=5cm]{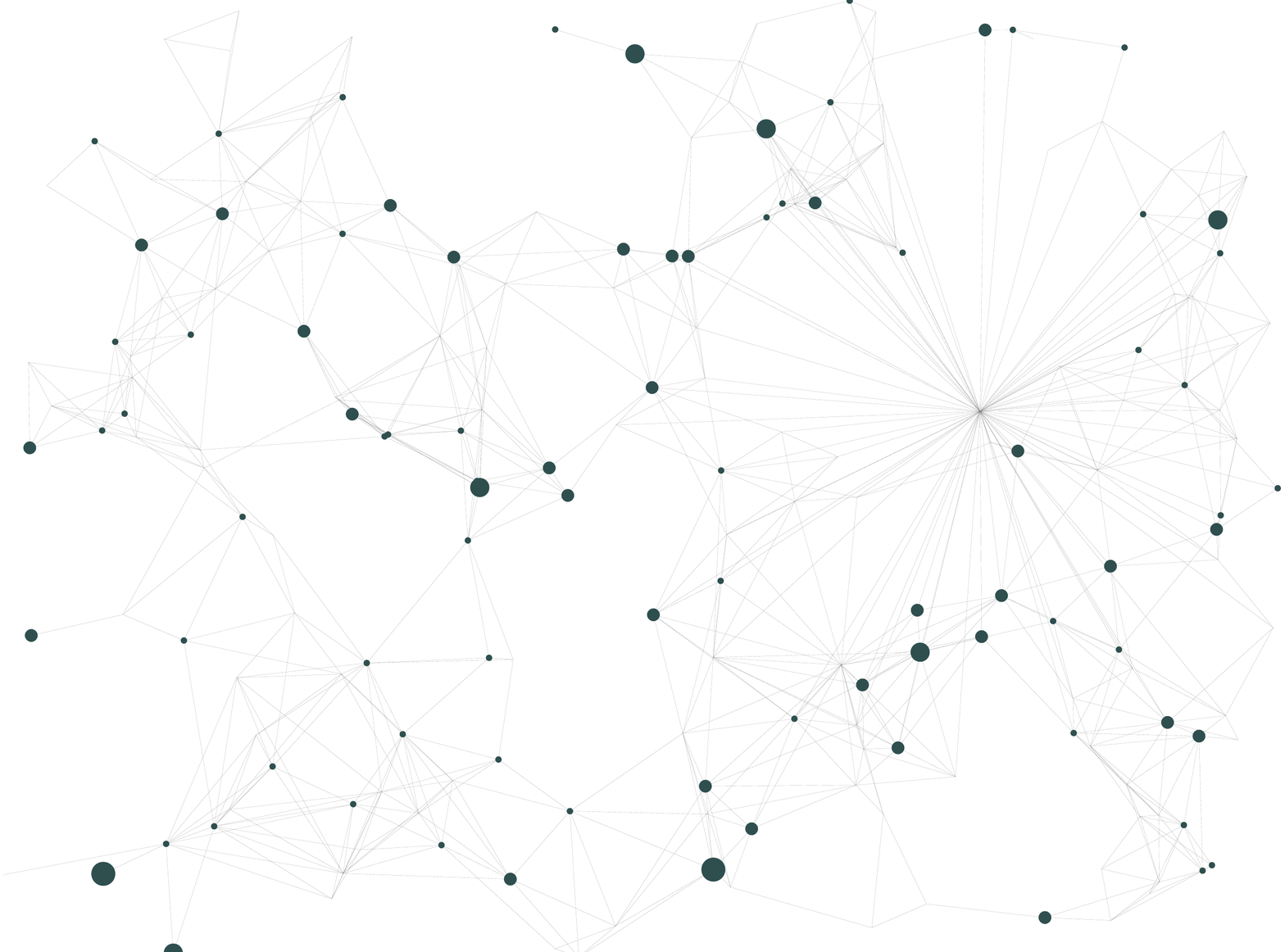}}
  \subfigure[Intermediate state]{\label{fig:Annealing:e} \includegraphics[width=5cm]{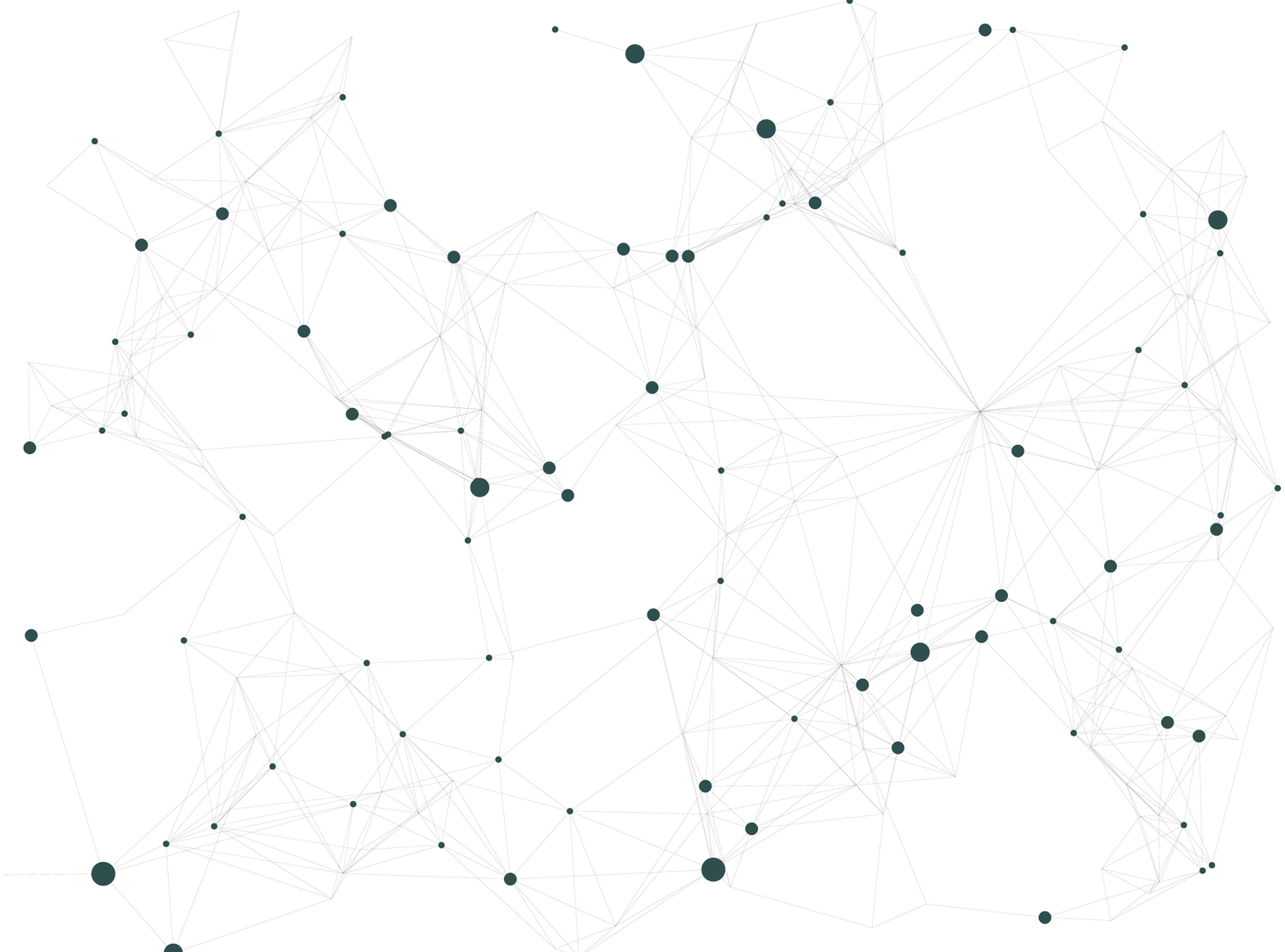}} 
  \subfigure[Equilibrium network]{\label{fig:Annealing:f} \includegraphics[width=5cm]{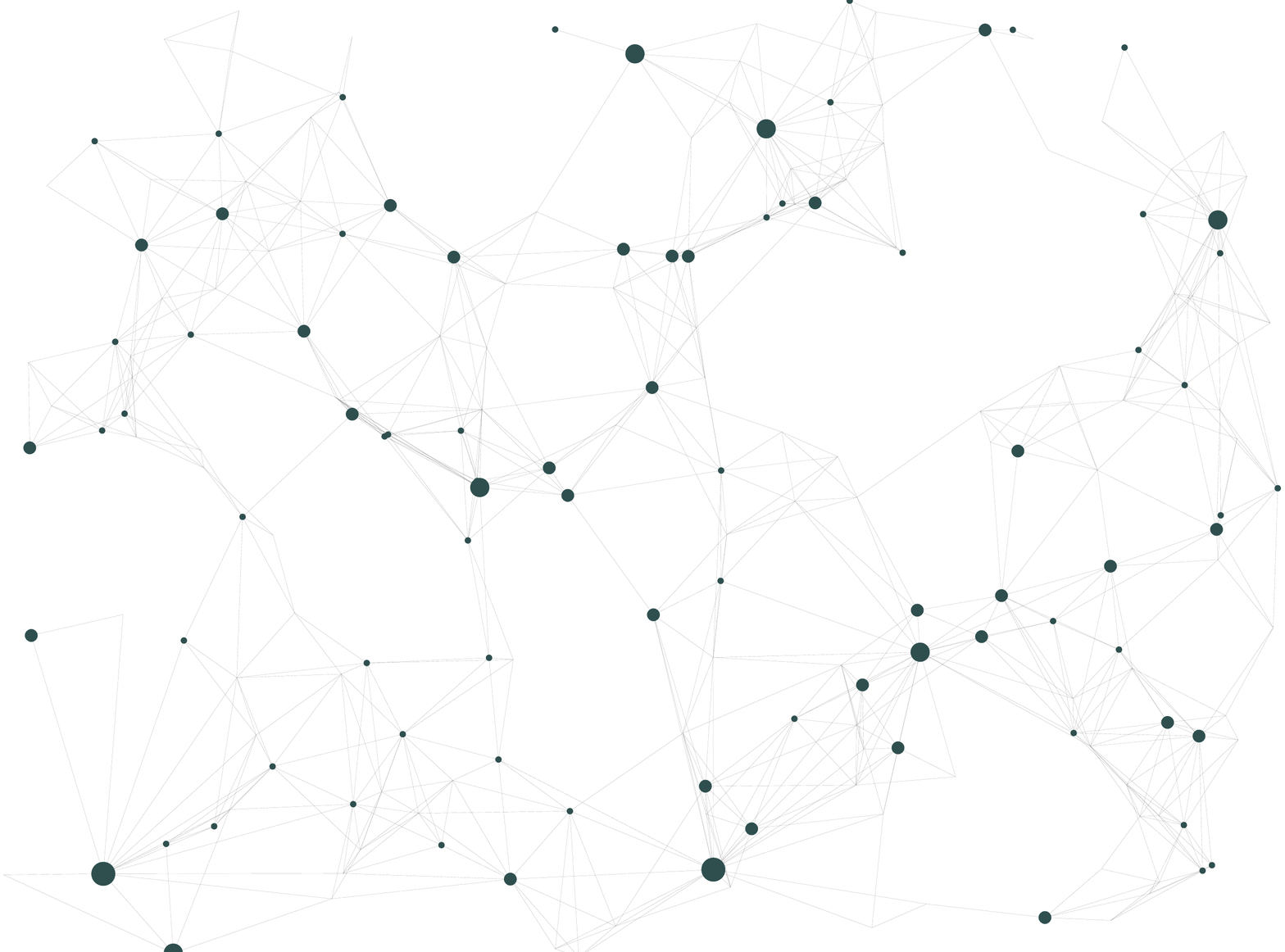}}
  \caption{\label{fig:Annealing}Equilibration of overlay topology embedded in Euclidean space based on a distributed annealing mechanism and a Pareto distribution of node resources (visualized by log-scaled node sizes)}
\end{figure}

The equilibrium perspective on structure formation phenomena in networks can be further applied in a straight-forward fashion in the development of \emph{distributed approaches to a run-time simulated annealing of overlays}. The result of such a mechanism is depicted in Figure \ref{fig:Annealing}. In this --admittedly simple-- illustrating example, nodes are assumed to be embedded into a virtual, two-dimensional Euclidean space. In practice, these could be geographic locations of devices in a ubiquitous computing scenario. Addressing a Peer-to-Peer setting, one may also think about a coordinate space created by distributed protocols like the Vivaldi latency estimation scheme \cite{Dabek2004}. We assume that nodes have time-varying bandwidth capacities that are distributed according to a heavy-tailed distribution and which are visualized in Figure \ref{fig:Annealing}, where node sizes are scaled logarithmically. Each connection between nodes further consumes a fixed amount of the bandwidth resources from both nodes.

For this simple model of a distributed system, the formation of structure in the overlay that is shown in Figure \ref{fig:Annealing} is due to a straight-forward application of the statistical mechanics generalization of the $G(n,p)$ random graph model. Here, the probability for a particular connection between nodes $v$ and $w$ to be constructed is assumed to be
\begin{equation*}
p(v,w) = \frac{1}{1+\hbox{e}^{(e(v,w)-\mu(v,w))/{T}}}
\end{equation*}
where $e(v,w)$ is the energy or cost of a connection between $v$ and $w$, $\mu(v,w)$ is the equivalent to a chemical potential in statistical mechanics and $T$ is a parameter controlling thermal fluctuations.\footnote{Here the reader may note that in the infinite temperature limit $T\rightarrow \infty$, this is equivalent to the $G(n,p)$ model for the special case $p=\frac{1}{2}$ in which all networks are equiprobable (independent of the number of edges).} In the simple illustrative example, the energy $e(v,w)$ of a connection is defined as a weighted combination of bandwidth demand and the squared Euclidean distance between $v$ and $w$. As chemical potential $\mu(v,w)$ we simply define the sum of bandwidth capacities of nodes $v$ and $w$. 

Thermal fluctuations of connections are due to a simple protocol: Each node $v$ periodically sends information (network address, bandwidth capacity and demand) about one random one-hop neighbor $w_1$ to another random one-hop neighbor $w_2$. If $w_1$ and $w_2$ are not connected, node $w_2$ calculates the probability with which it rewires the link pointing to $v$ into $w_1$, based on the equation given above. The result of applying this scheme to an initial random network is the formation of a topology in which the connectivity is heuristically optimized with respect to Euclidean distance and capacity/demand constraints (see Figures \ref{fig:Annealing:a} - \ref{fig:Annealing:c}). When the capacities of nodes change (see Figure \ref{fig:Annealing:d}), the topology adapts accordingly just due to thermal fluctuations (Figures \ref{fig:Annealing:e} - \ref{fig:Annealing:f}).

While one needs to be aware that the feasibility of such a simple annealing approach in practice critically depends on the ruggedness of the ``energy landscape'' and the overhead induced by thermal fluctuations, this example is meant to illustrate the equilibrium statistical mechanics perspective on \emph{organic network structures}. The formation of optimized structures in the above example is a direct application of crystallization phenomena in the natural world. In real-world settings, one can further imagine rewiring schemes that make use of the natural random fluctuations of nodes in overlay networks, thus reducing the overhead of active rewiring operations.

\subsection{Enforcing Ensembles - The Micro-Macro Link}
\label{sec:Organic:MicroMacro}

In section \ref{sec:Networks:Complex}, we have commented on the rich body of results on collective network properties like connectedness, diameter, resilience against faults and attacks, as well as on the performance of dynamical processes like information dissemination, synchronization, consensus and gossiping schemes. All these statements are based on a reasoning in terms of statistical ensembles, so the guarantees that one can give for an actual overlay topology are necessarily stochastic. However, statements on properties that hold asymptotically almost surely become more reliable as the size of the network topology (in terms of the number of nodes) increases, just like statements on properties of thermodynamic systems become more reliable as the volume being considered is increased. Relying on statements for instance about the diameter or connectedness of topologies emerging from a stochastic process becomes more and more reasonable as networked computing systems are growing in size\footnote{Since in practice one necessarily deals with finite-size systems, it is important to note that technically one also needs to consider how fast the associated probability converges to $1$ as the network size increases. It is often possible to give analytical expressions for these so-called finite-size effects. While we refer the interested reader to \cite{Dorogovtsev2003} for more details, at this point it is sufficient to note that the properties mentioned above hold for networks whose size is reasonable small (in the order of a few hundred to a few thousand nodes) to be of practical value for the envisioned scenario.}. While being based on a different conceptual framework, the idea of using stochastic reasoning to facilitate the design of massively distributed systems with unreliable components resembles the notion of \emph{thermodynamic system design} that has been proposed in \cite{Kubiatowicz2003} and considered more recently in \cite{Parunak2011}. The most interesting aspect of this approach is that it actually becomes --given that they adhere to a particular statistical ensemble-- easier to make substantiated statements about networked systems as they are becoming larger. 

So far, we have discussed the link between the abstraction of constrained adiabatic ensembles of networks and collective network properties. However, from a practical perspective this link is useless if we are not able to relate the ensemble description of a system with an actual overlay topology resulting from a distributed topology management scheme. In real-world systems, one may be confronted with situations in which the actual stochastic dynamics driving the topology construction is given or influenced by some external factors that cannot be influenced. In such situations one can use the set of tools of statistical mechanics (like e.g.~master equations or mean-field approximations) to directly derive aggregate statistical quantities of interest --and thus the ensemble description of the system-- from a stochastic description of the dynamics of individual nodes (see e.g. the application of this procedure in \cite{Barabasi1999b} and subsequent refinements to describe the evolution of the topology of the Gnutella Peer-to-Peer system). In many cases, the situation is however different, as we may wish to actually design a topology management scheme that efficiently constructs a network topology (drawn from a particular constrained adiabatic ensemble) which has desirable properties. In this case, one can employ a \emph{configuration approach} to directly derive the local dynamics from the ensemble description. In the following we will demonstrate this approach for the particular case of random scale-free networks. 

\begin{figure}[!ht]
  \centering
  \includegraphics[width=6cm]{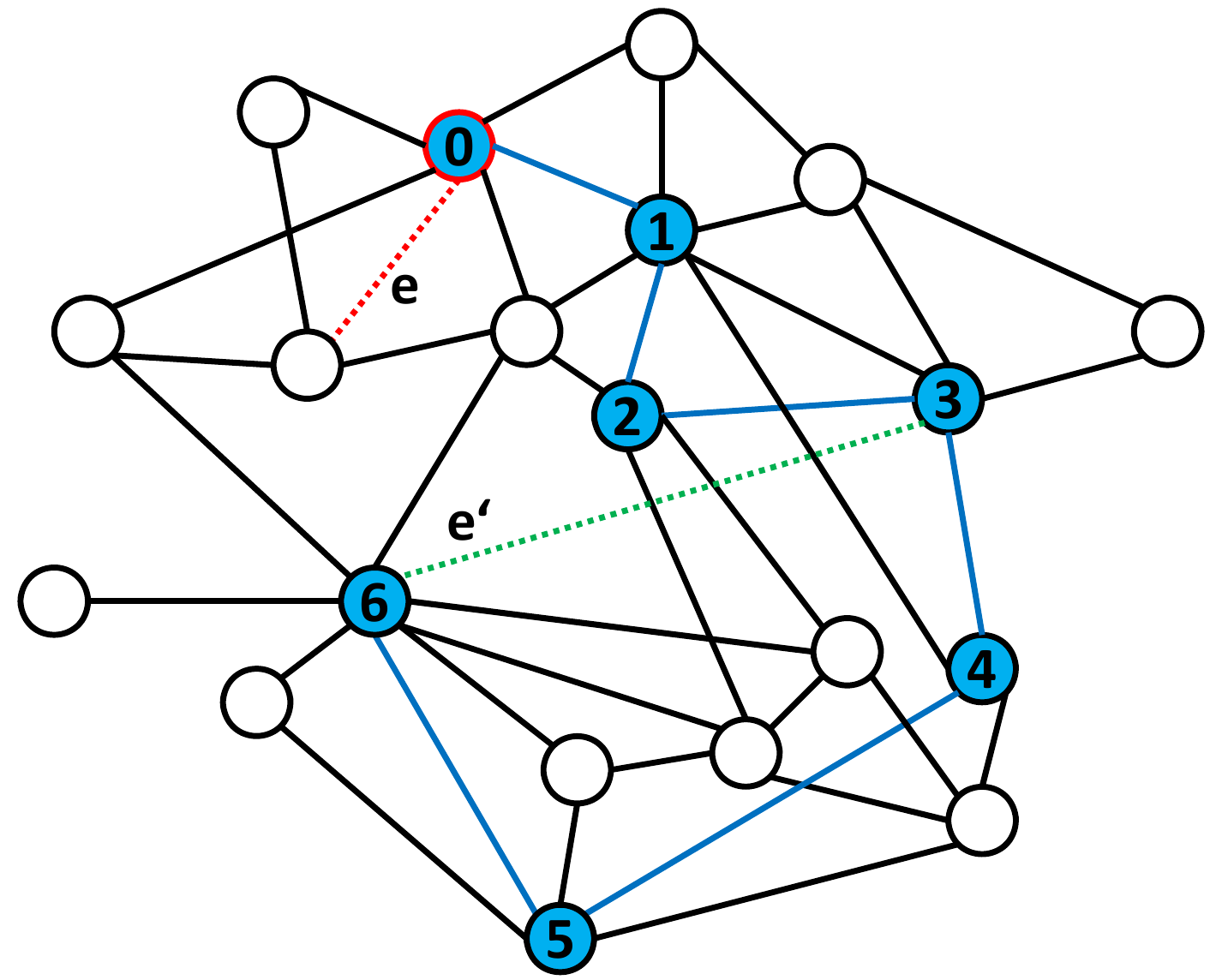}
  \caption{\label{fig:Rewiring} Rewiring of edge $e$ initiated by node $0$ by means of two consecutive random walks with step length three}
\end{figure}

The actual distributed mechanism has been presented and evaluated in much detail in \cite{Scholtes2010,Scholtes2010a}. The basic idea is to start with an arbitrary connected topology that has been generated for instance by a bootstrapping process. One then progressively rewires all connections by means of a biased random walk sampling scheme like for instance the one proposed in \cite{Zhong2008}. Again referring to the aforementioned articles for  a more detailed algorithmic description of the protocol, each rewiring of an existing connection $e$ between nodes $i$ and $j$ is initiated by one of their endpoints. This node deletes the connection $e$ and creates a sampling message that contains the addresses of $i$ and $j$. By means of two consecutive biased random walks of length $l$, two new nodes $v$ and $w$ are then sampled which will form a new connection $e'$ if it does not already exist. This process is illustrated in Figure \ref{fig:Rewiring}, which shows a random walk rewiring of edge $e$ initiated by node $0$. A first random walk takes three steps and finds node $3$. After another $3$ steps the second endpoint --node $6$-- is found and the new edge $e'$ is created. Similar to the illustrative example in section \ref{sec:Organic:Equilibrium}, by this scheme one effectively models random particle motion based on an artificially designed energy landscape that constrains the accessible configurations. However, in this case, our starting point is a particularly constrained --and possibly fine-tuned-- statistical ensemble which we wish to enforce rather than a local definition of energies and chemical potentials. 

\begin{figure}[!ht]
  \centering
  \subfigure[Average fitted exponent $\gamma_f$]{\label{fig:ER:Gamma} \includegraphics[width=5.5cm]{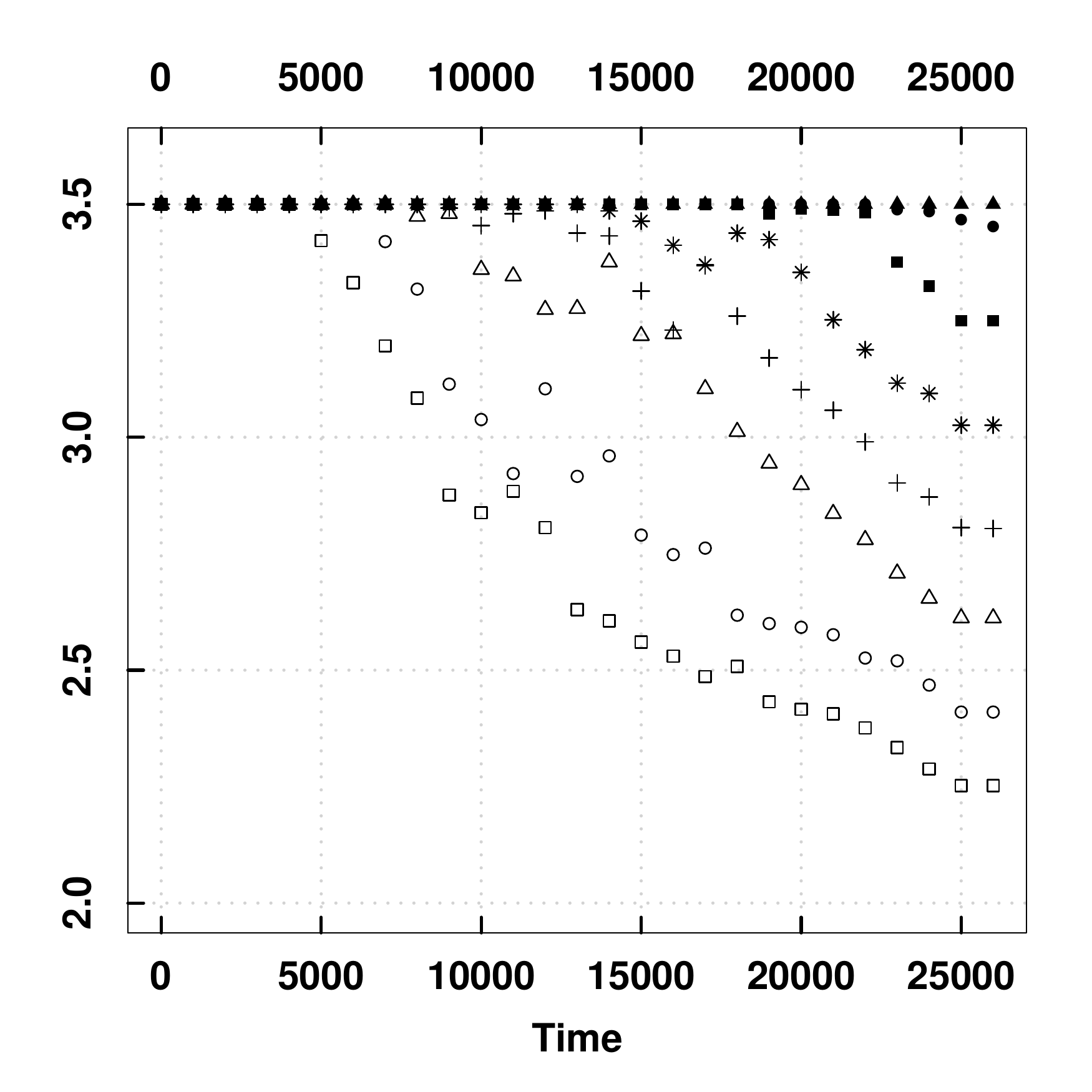}}		
  \subfigure[Average Kolmogorov-Smirnov statistic $D$]{\label{fig:ER:D} \includegraphics[width=5.5cm]{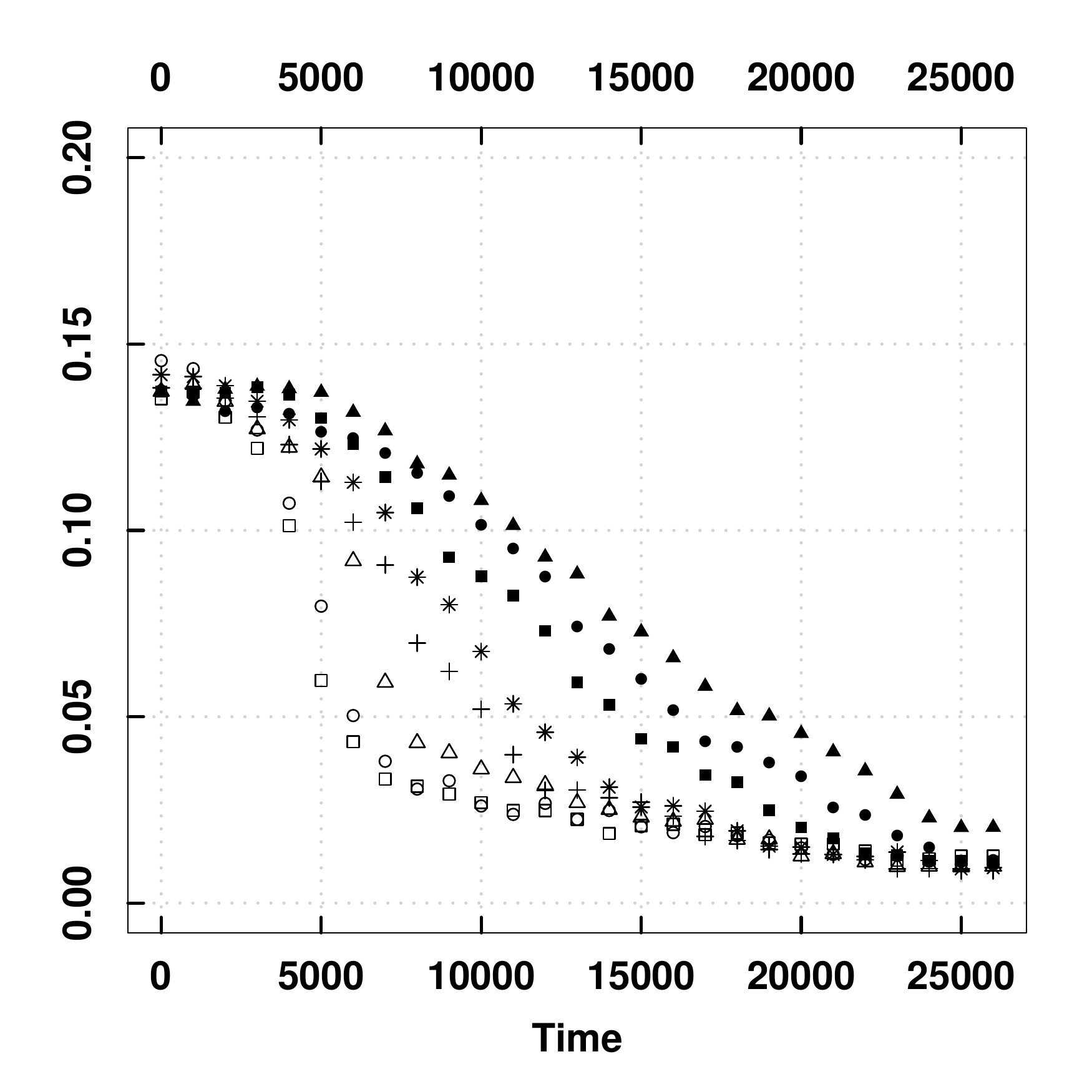}} \qquad
  \caption{\label{fig:ER} Time evolution of 5000 node networks during adaptation runs with $\gamma \in [2.1, 3.5]$}
\end{figure}

Given that the transition kernel of the biased random walk sampling is configured appropriately and the number of steps taken by the random walk is sufficiently large, the topology is guaranteed to be drawn from a given constrained adiabatic ensemble once all connections have been rewired. Numerically we have found for the required random walk length a maximum of $\log(n)$ steps to be sufficient where $n$ is the number of nodes. An upper bound for the required random walk length, which is however not tight, can actually be derived based on the framework described in \ref{sec:Organic:Dynamic} (see details in \cite{Scholtes2010}. Based on a configuration model approach for an adiabatic ensemble of scale-free networks that has been introduced in \cite{Lee2004a} as well as the Metropolis-Hastings algorithm \cite{Hastings1970} we can actually derive the required transition kernel for the random walk. In order to effectuate an ensemble with a particular exponent $\gamma$ one has to configure the random walk such that each node $i$ forwards a sampling message to a neighbor $j$ with probability
\begin{equation}
  \label{equ:TransitionProb}
  P_{i,j} = \frac{d_i}{d_j} \left( \frac{i}{j} \right)^{\frac{1}{\gamma-1}}
\end{equation}
where $i,j$ are numeric, not necessarily unique node identifiers chosen uniformly at random from an identifier space. The effect of the resulting rewiring process on the degree distribution is shown in Figure \ref{fig:ER} in terms of the evolution of fitted degree distribution exponent as well as the Kolmogorov-Smirnov statistic $D$ which quantifies the goodness of the fit. The decreasing values for $D$ show that the hypothesis that the topology is indeed a random network with a power-law degree distribution becomes more reasonable. Since the network is sampled at random, for this simple topology management mechanism one can thus safely rely on all analytical results that hold for the ensemble of random scale-free networks with the chosen exponent.

\subsection{Adaptation in Organic Overlays - Triggering Phase Transitions}
\label{sec:Organic:Transitions}

\begin{figure}[!ht]
  \centering
  \includegraphics[width=8cm]{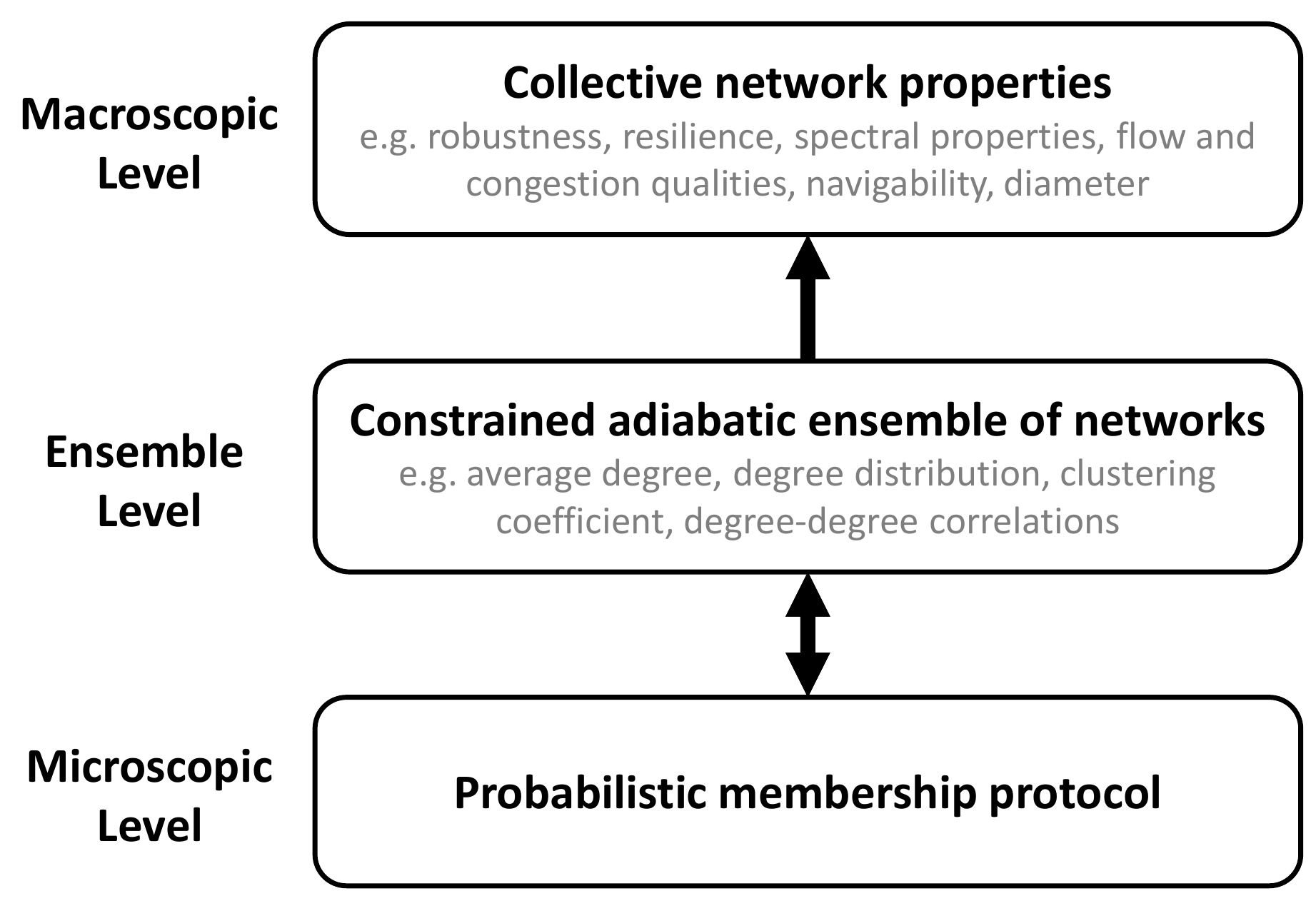}
  \caption{Micro-macro link provided by a thermodynamic approach to the design of overlay networks}\label{fig:micromacro}
\end{figure}

The intriguing aspect of the results above is that this procedure provides an analytical link between a stochastic model for the micro-scale dynamics of a distributed system and collective properties that are observable at the system level. 
Furthermore, this is independent on what approach is chosen to relate the microscopic dynamics (at the level of individual nodes and protocol messages) with an ensemble description of the overlay topology.
The links between these three levels of description is depicted in Figure \ref{fig:micromacro}. Systems with adaptive thermodynamic structures can be built based on this perspective when considering --at the intermediate ensemble level-- a discrete set or continuum of ensembles with different constraints and enforcing a particular ensemble based on the environmental conditions.

In this sense, the exponent parameter $\gamma$ that influences the random walk bias in the forwarding of sampling messages, effectively provides a control knob that allows to switch between different points in a continuum of different constrained adiabatic ensembles giving rise to potentially different macroscopic properties. While this micro-macro link and the possibility to change macroscopic properties by tuning a sampling bias is useful \textit{per se}, a further interesting aspect are phase transition phenomena, i.e. the existence of critical points in the space of control parameters at which the properties of the resulting topology change suddenly. This may involve smooth ({\em second-order}) or abrupt ({\em first-order}) changes as well as phase transitions with hysteresis effects. Such critical points are of primary interest for the complex networks community and a number of analytical results have been obtained that relate these points to a qualitative change of aggregate statistics. A particularly well-studied type of networks are those with degree distributions that follow a power-law. This means that, in the limit of infinite-size networks, the probability for a node to have exactly $k$ connections is given by a Zeta (also called {\em scale-free}) distribution,
\begin{equation*}
  P(k) = \frac{k^{-\gamma}}{\zeta(\gamma)}
\end{equation*}			
with $\zeta:\mathbb{R}\rightarrow\mathbb{R}$ being the real-valued Riemann zeta function, 
$  \zeta(\gamma) = \sum_{i=1}^{\infty}{i^{-\gamma}}$.
Many results about critical phenomena in networks with fixed degree distributions are due to the so-called Molloy-Reed criterion \cite{Molloy1995} which links the relation of the distribution's first two moments to the existence of a giant connected component. This has been also been used successfully to study --as mentioned already before-- the error and attack tolerance of random power-law networks. For this special case, in \cite{Cohen2000} it was found that at least a fraction
\begin{equation*}
 r := 1 - \left( \frac{\zeta(\gamma-2)}{\zeta(\gamma-1)} -1 \right)^{-1}
\end{equation*}
of nodes need to fail at random for the giant connected component of a power-law network to be destroyed. Regarding the control parameter $\gamma$ --which determines the behavior of the distributed rewiring scheme described in section \ref{sec:Organic:MicroMacro}-- the convergence behavior of the Zeta function results in a phase transition once the parameter crosses the critical point $\gamma=3$. For $\gamma\geq 3$, the terms $\zeta(\gamma-2)$ and $\zeta(\gamma-1)$ are constants, thus resulting in a constant non-zero value for the critical fraction $r$. For $\gamma\in(2,3)$, the term $\zeta(\gamma-2)$ diverges. In this range $r \rightarrow 1$, i.e. almost all nodes can be removed at random without destroying the giant connected component. The parameter $\gamma$ thus effectively allows to produce a continuum of constrained adiabatic ensembles while at the critical point $\gamma=3$,  the resilience properties of the resulting networks undergo a qualitative change.

\begin{figure}[!ht]
  \centering
  \subfigure[Power Law Exponent $\gamma$]{\label{fig:PhaseTrans:Gamma} \includegraphics[width=5.5cm]{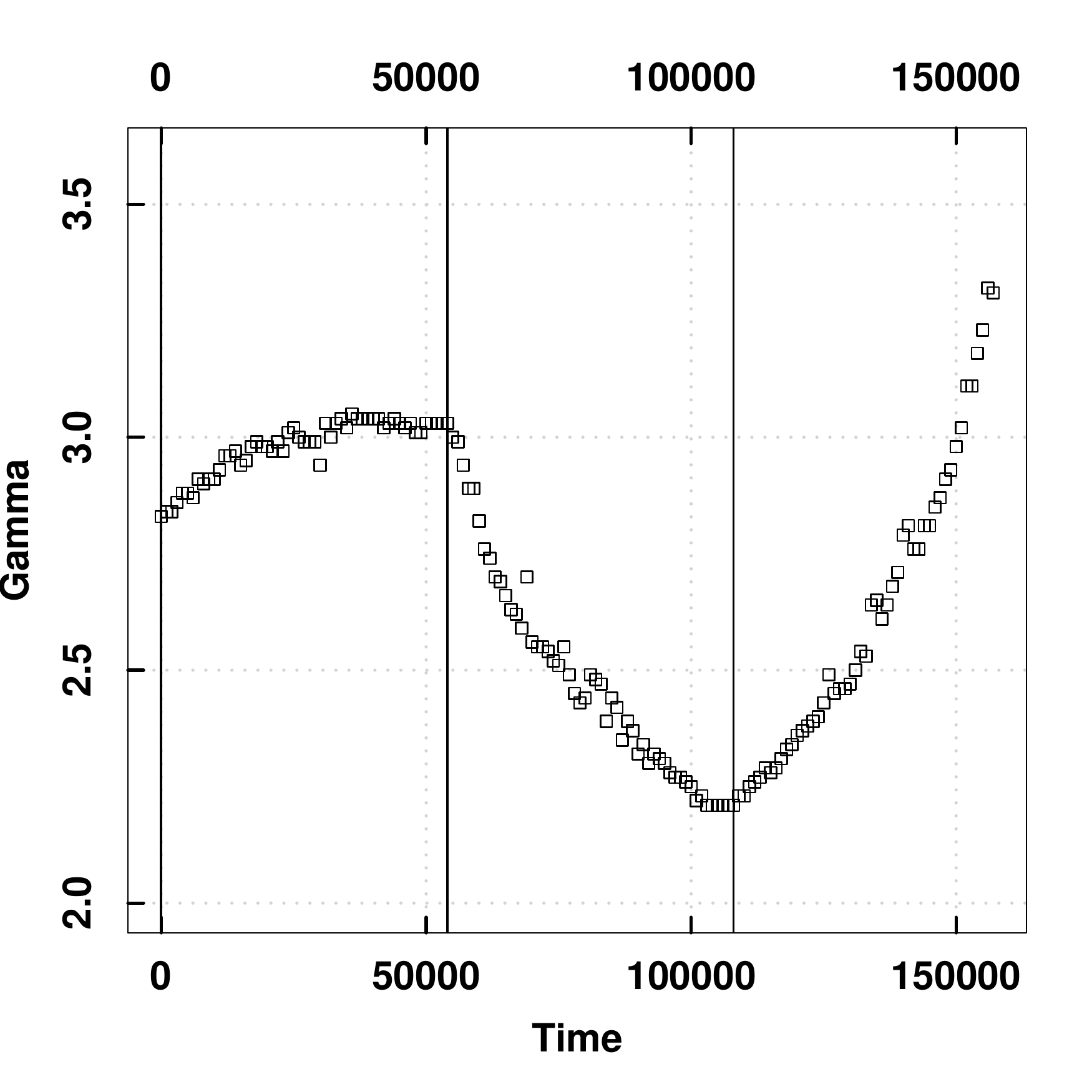}}
  \subfigure[Kolmogorov-Smirnov statistic $D$]{\label{fig:PhaseTrans:D} \includegraphics[width=5.5cm]{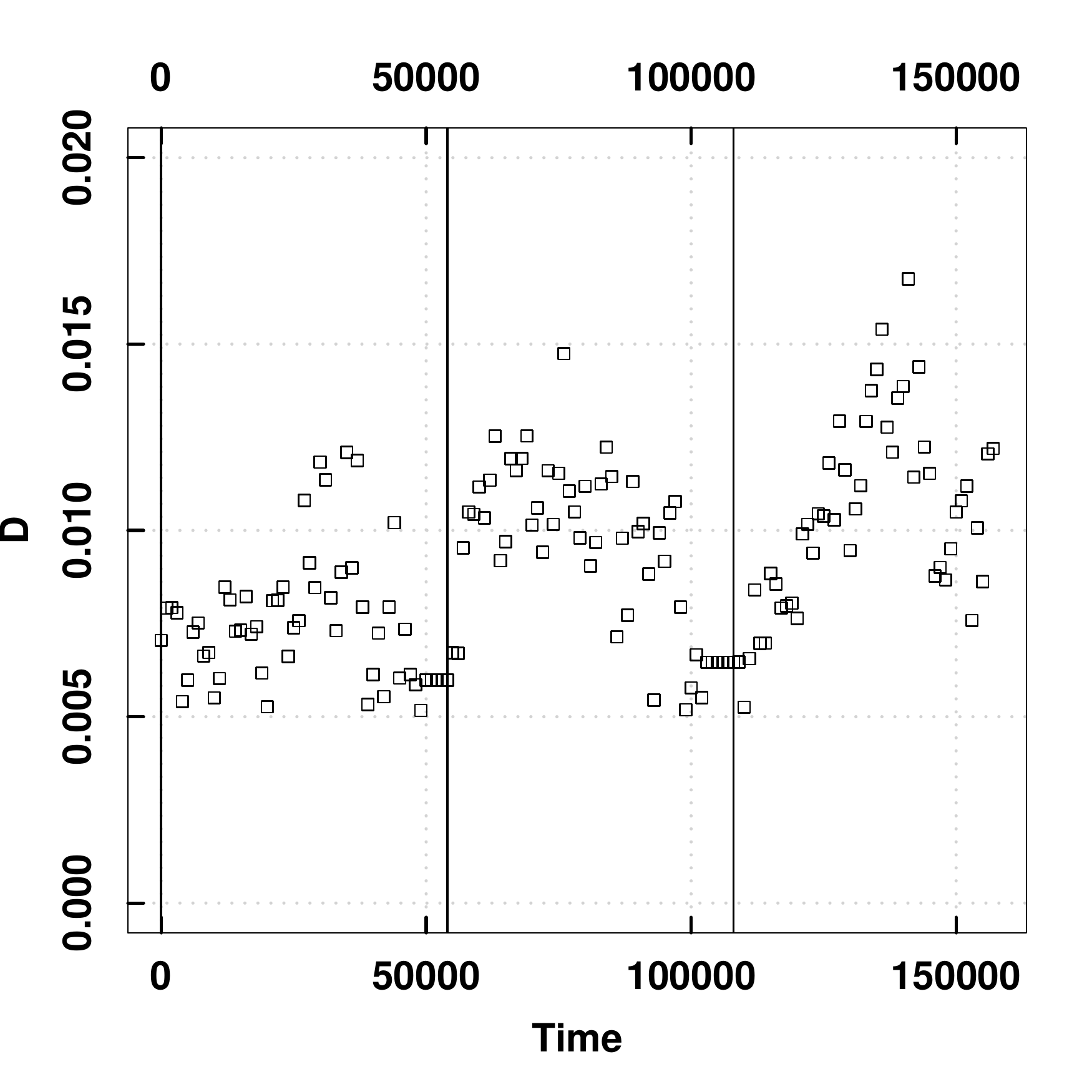}} \qquad 
  \subfigure[Result of attack for network with $\gamma \approx 2$]{\label{fig:PhaseTrans:Attack:a} \includegraphics[width=5.5cm]{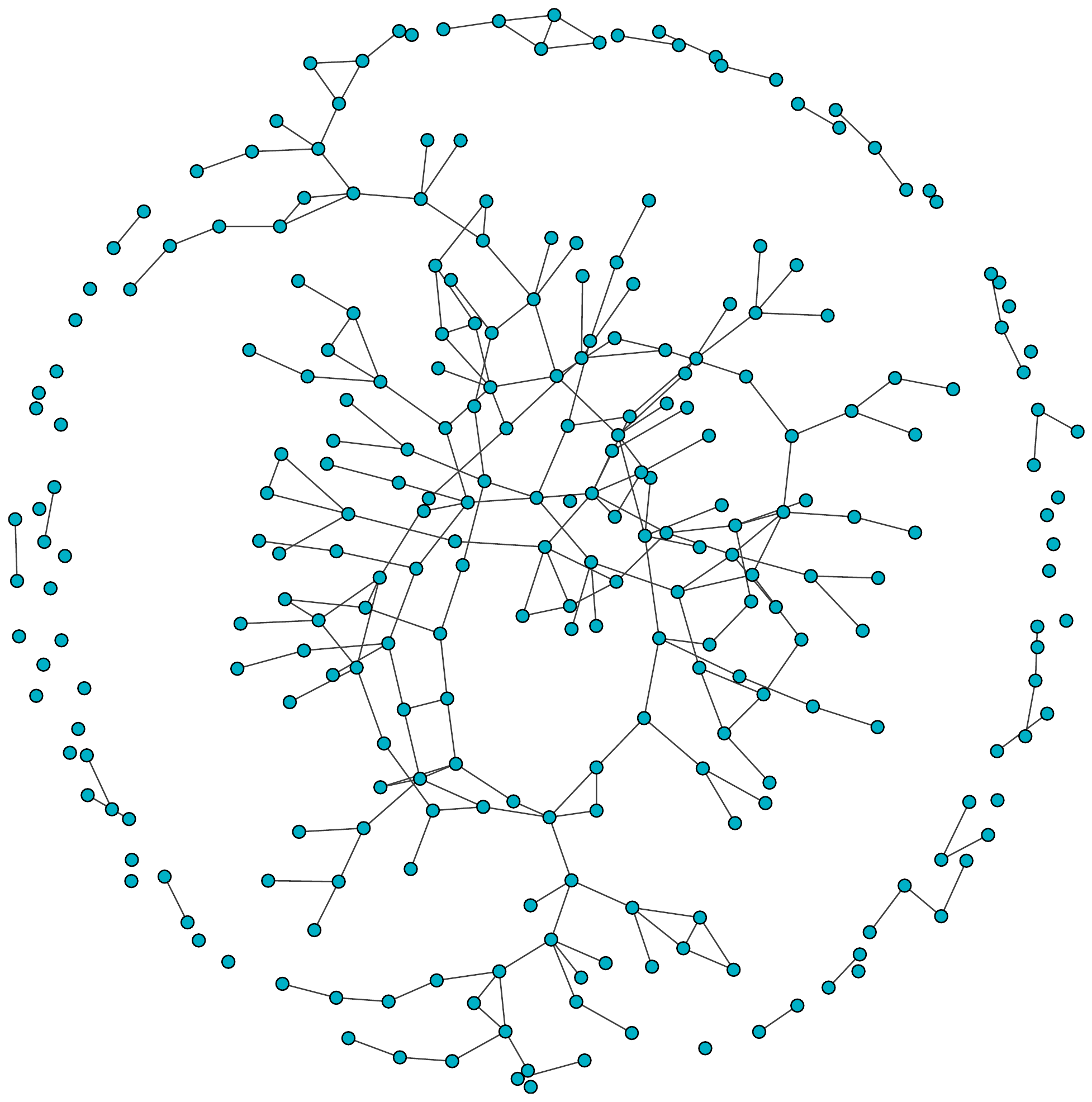}}
  \subfigure[Result of attack for network with $\gamma \approx 3.5$ ]{\label{fig:PhaseTrans:Attack:b} \includegraphics[width=5.5cm]{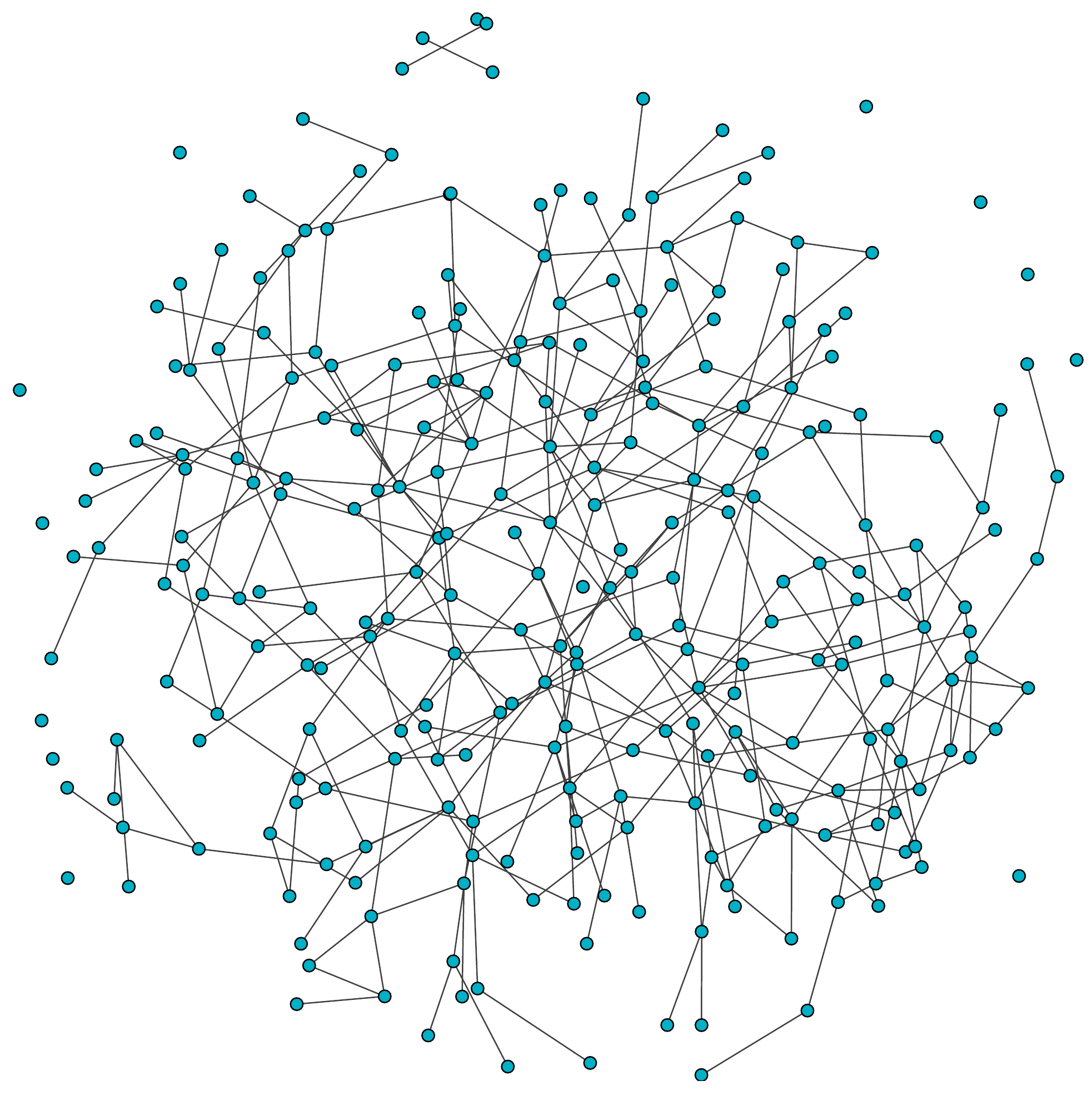}}
\caption{\label{fig:PhaseTrans}Time Evolution of scale-free network during multiple adaptation cycles. Start/end times of adaptation cycles are indicated by vertical lines.}
\end{figure}

The same argument about the change in the convergence behavior of the Zeta function has actually been used to derive phase transitions in terms of diameter, attack resilience \cite{Cohen2001} or the efficiency of spreading phenomena \cite{Moreno2004}. Given that --depending on the used distributed algorithms, as well as the current operational conditions-- these properties can be both desirable or detrimental. So, the knowledge about these effects can actually be used for an active adaptation of collective network qualities. For the random walk rewiring described in section \ref{sec:Organic:MicroMacro}, all one has to do is to change the control parameter $\gamma$ in equation \ref{equ:TransitionProb}. Figure \ref{fig:PhaseTrans} shows simulation results that have been obtained based on this idea and that have  been presented originally in \cite{Scholtes2010}. Here, the random rewiring protocol described above has been applied continuously, while at certain times (indicated by vertical lines in the Figures \ref{fig:PhaseTrans:Gamma} and \ref{fig:PhaseTrans:Gamma}) the parameter $\gamma$ was changed. After each modification, the network was allowed to equilibrate by progressively resampling all connections according to the new constrained adiabatic ensemble. Figure \ref{fig:PhaseTrans:Gamma} shows the evolution of the fitted degree distribution exponent $\gamma$. The Kolmogorov-Smirnov statistic $D$ depicted in Figure \ref{fig:PhaseTrans:D} shows the goodness of the assumption that the connectivity distribution indeed follows a power-law with the fitted exponent. Smaller values for $D$ again represent a larger reliability of the fit. Here, it can clearly be seen that near the end of the adaptation cycles (which was at the same time the beginning of a new one with a different $\gamma$), the degree distribution of the network does indeed follow a power-law.

Based on the theoretical findings regarding the critical point of $\gamma=3$ for the robustness of the topology against attacks, one would expect that the robustness changes qualitatively between the last two adaptation rounds, i.e. when the exponent of the topology is increased from $2.1$ to $3.5$. To underpin this theoretical result, Figures \ref{fig:PhaseTrans:Attack:a} and \ref{fig:PhaseTrans:Attack:a} show the network topology that remains after $10 \%$ of the most connected nodes have been removed from a $300$ node network at these two different points in the adaptation process. The large number of isolated nodes and clusters in Figure \ref{fig:PhaseTrans:Attack:a} and the comparison with the topology shown in \ref{fig:PhaseTrans:Attack:b} clearly show the practical implications of this theoretical finding in terms of resilience. In summary, this particular phase transition effect in equilibrium ensembles of scale-free networks can be used to make a trade-off between desirable and detrimental properties. In phases where efficient spreading is needed, the topology can be sampled from an adiabatic ensemble with $\gamma \in (2,3)$. Similarly, according to \cite{Sarshar2004} a $\gamma \in(2,2.3)$ should be chosen to maximize the efficiency of a random walk based distributed search in scale-free topologies. In situations where attacks or a spreading of failures are being detected, the local connection sampling can be instrumented such that the resulting topology is much more resilient against these effects, while at the same time reducing the efficiency of distributed search and information dissemination.

\newpage
\section{Conclusion and Outlook}
\label{sec:Conclusion}

In this article, we have summarized the statistical mechanics and the statistical physics perspective on the modeling of complex network structures. We have then outlined some ideas on how this perspective can be used constructively in the management of overlay networks for very large, dynamic systems. In particular, we argue that --at least in very large and highly dynamic systems-- it can be easier to enforce a particular statistical ensemble which will give rise to desirable macro-level properties and performance of distributed schemes than using sophisticated topology maintenance schemes. By means of active randomization of protocols, one can thus obtain strong, \emph{thermodynamic} guarantees for the structure and dynamics emerging in sufficiently large systems. For the design of robust and adaptive organic computing systems, we thus argue that randomization and \emph{loosening precise control} are crucial ingredients.


While we have showed first examples of actual distributed mechanisms that allow to rely on the intriguing results of complex network science, this work is necessarily incomplete. For the future, we envision for instance novel mechanisms that make use of the natural dynamics of Peer-to-Peer systems in the efficient construction and adaptation of an overlay topology with complex, yet predictable structures and properties. For this, rather than actively rewiring connections, one can use the natural turnover of machines and users (usually called churn) and apply random connection sampling schemes only as nodes join the system, change their characteristics or exhibit communication errors that lead to connections being removed. By assuming that the system is in equilibrium, the collective properties of the overlay network can then be predicted in analogy to the analysis of many-particle systems by means of statistical mechanics. Moreover, under this framework also the costs induced by structure maintenance protocols are massively reduced.

Referring again to the scenario depicted in section \ref{sec:Introduction}, one may thus be tempted to summarize the challenges of future systems, as well as the idea of addressing them in the framework presented in this article in the following way:

\begin{quote}
\textit{As network devices become more akin to particles in terms of number, size and stochastic behavior, can we design distributed systems along models, methods and abstractions from statistical mechanics and thermodynamics?}
\end{quote}

Considering recent advances in the statistical physics' study of complex systems in general, and complex networks in particular, one may argue that this is true at least for some aspects of large dynamic systems. It is thus reasonable to foresee a number of potential applications of complex network science in the engineering of organic computing systems and in the design of mechanisms for sustainable future techno-social systems.

\bibliographystyle{unsrt}
\bibliography{library}

\end{document}